\documentclass[%
 aip,
 amsmath,amssymb,
 reprint,%
]{revtex4-1}

\usepackage{graphicx}
\usepackage{dcolumn}
\usepackage{bm}

\usepackage[utf8]{inputenc}
\usepackage[T1]{fontenc}
\usepackage{mathptmx}
\usepackage{etoolbox}

\makeatletter
\def\@email#1#2{%
 \endgroup
 \patchcmd{\titleblock@produce}
  {\frontmatter@RRAPformat}
  {\frontmatter@RRAPformat{\produce@RRAP{*#1\href{mailto:#2}{#2}}}\frontmatter@RRAPformat}
  {}{}
}%
\makeatother
\begin{document}

\preprint{AIP/123-QED}

\title[Polycrystalline MnBi as a transverse thermoelectric material]{Polycrystalline MnBi as a transverse thermoelectric material}
\author{A. Sola}
 \email{a.sola@inrim.it}
\affiliation{ 
	Istituto Nazionale di Ricerca Metrologica, Strada delle Cacce 91, 10135, Turin, Italy.
}%
\author{E. Olivetti}%
\affiliation{ 
	Istituto Nazionale di Ricerca Metrologica, Strada delle Cacce 91, 10135, Turin, Italy.
}%

\author{L. Martino}
\affiliation{ 
	Istituto Nazionale di Ricerca Metrologica, Strada delle Cacce 91, 10135, Turin, Italy.
}%

\author{V. Basso}
\affiliation{ 
	Istituto Nazionale di Ricerca Metrologica, Strada delle Cacce 91, 10135, Turin, Italy.
}%

\date{\today}

\begin{abstract}
To assess the potential of polycrystalline MnBi as a transverse thermoelectric material, we have experimentally investigated its anomalous Nernst effect (ANE) by means of the heat flux method. We prepared MnBi samples by powder metallurgy; this technique allows the preparation of samples in arbitrary shapes with the possibility to tailor their magnetic properties. In the material exhibiting the highest remanent magnetization, we found a value of the ANE thermopower of -1.1 $\mu$V/K at 1 T, after the compensation of the ordinary Nernst effect from pure bismuth present inside the polycrystalline sample. This value is comparable with those reported in the literature for single crystals. 
\end{abstract}

\maketitle

The performance of a thermoelectric device working with the Seebeck/Peltier effects is ruled by both its architecture and the characteristics of the materials. The latter are represented by the dimensionless figure of merit $ZT=\epsilon^{2}\sigma T/k$ where $\epsilon$ is the Seebeck termopower, $\sigma$ is the electrical conductivity, $T$ is the temperature and $k$ is the thermal conductivity. 
The typical device architecture consists in pillars of p-type and n-type semiconductors that are thermally in parallel and electrically in series \cite{disalvo1999thermoelectric}.
This configuration allows the maximization of the efficiency, but limits the possibilities of arbitrary shaping.
Structures such as thermal-sensing coatings \cite{kirihara2012spin} or flexible devices \cite{kirihara2016flexible} can however be devised by exploiting transverse thermoelectric effects \cite{boona2021transverse}, also called thermomagnetic effects \cite{nolas2001thermoelectrics}.
The Nernst effect refers to the transverse thermopower $N = -\nabla_{\mathrm{y}}V/\nabla_{\mathrm{x}}T$ that rises in a material in which the temperature gradient $\nabla_{\mathrm{x}}T$ is applied in presence of a magnetic field $B_{\mathrm{z}}$. In ferromagnetic materials, with a spontaneous magnetization, the thermopower is called spontaneous or anomalous Nernst effect (ANE).
The performance of a thermomagnetic device depends on the ratio between two lengths: one along the electric potential and the other along the thermal gradient. These are oriented along perpendicular directions, while in a Seebeck/Peltier device the two lengths are parallel and coincide. 
This fact opens new possibilities for the improvement of the performance, along with more freedom in the geometries, by tuning dimensions and shapes of the devices. The optimum shape, for an architecture based on multistage cascading, was derived by Harman \cite{harman1963theory} and subsequently developed \cite{kooi1968thermoelectric,scholz1994infinite,polash2021infinite}.
The opportunity to adapt the shape of thermoelectric devices to the geometries of existing systems is of paramount importance and sometimes it is the only option, as for example for the proposal of self-cooling cables \cite{de2016thermomagnetic}. 
In this framework, the study of thermomagnetic effects in ferromagnets that could be easily produced in different sizes and shapes, like polycrystalline materials, becomes crucial.
A material worth of investigation is MnBi, which allows the preparation of magnetic samples by powder metallurgy processes.
The thermomagnetic properties of MnBi single crystals have been recently investigated \cite{he2021large}, with some interesting findings in terms of an extra contribution to the transverse thermopower. This originates from magnon-electron interaction due to the large spin-orbit coupling of the MnBi itself and it was also observed in heterogeneous systems such as multi-layers \cite{lee2015thermoelectric,ramos2019interface} and composites of bulk materials with heavy metal nanoparticles \cite{boona2016observation}. Nevertheless the possibility of exploiting this favorable condition in a single material makes the MnBi appealing in the framework of the thermomagnetic applications.
Previous studies on MnBi single-crystals report a transverse thermopower at saturating field of 2 $\mu$V/K at room temperature and 10 $\mu$V/K at 80 K \cite{he2021large}. These values are comparable with the maximum ANE thermopowers reported in a recent perspective paper by Uchida and coworkers \cite{uchida2021transverse}, where the highest ANE coefficients are reported for the Heusler ferromagnet Co$_2$MnGa.
Some experimental results of ANE in polycrystalline samples are already present in the literature, with large values at 2T recently reported for polycrystalline Fe$_{\mathrm{x}}$Ga$_{\mathrm{4-x}}$ (2.96 < x < 3.15) \cite{feng2022giant}.

\begin{figure}
\includegraphics[scale=0.4]{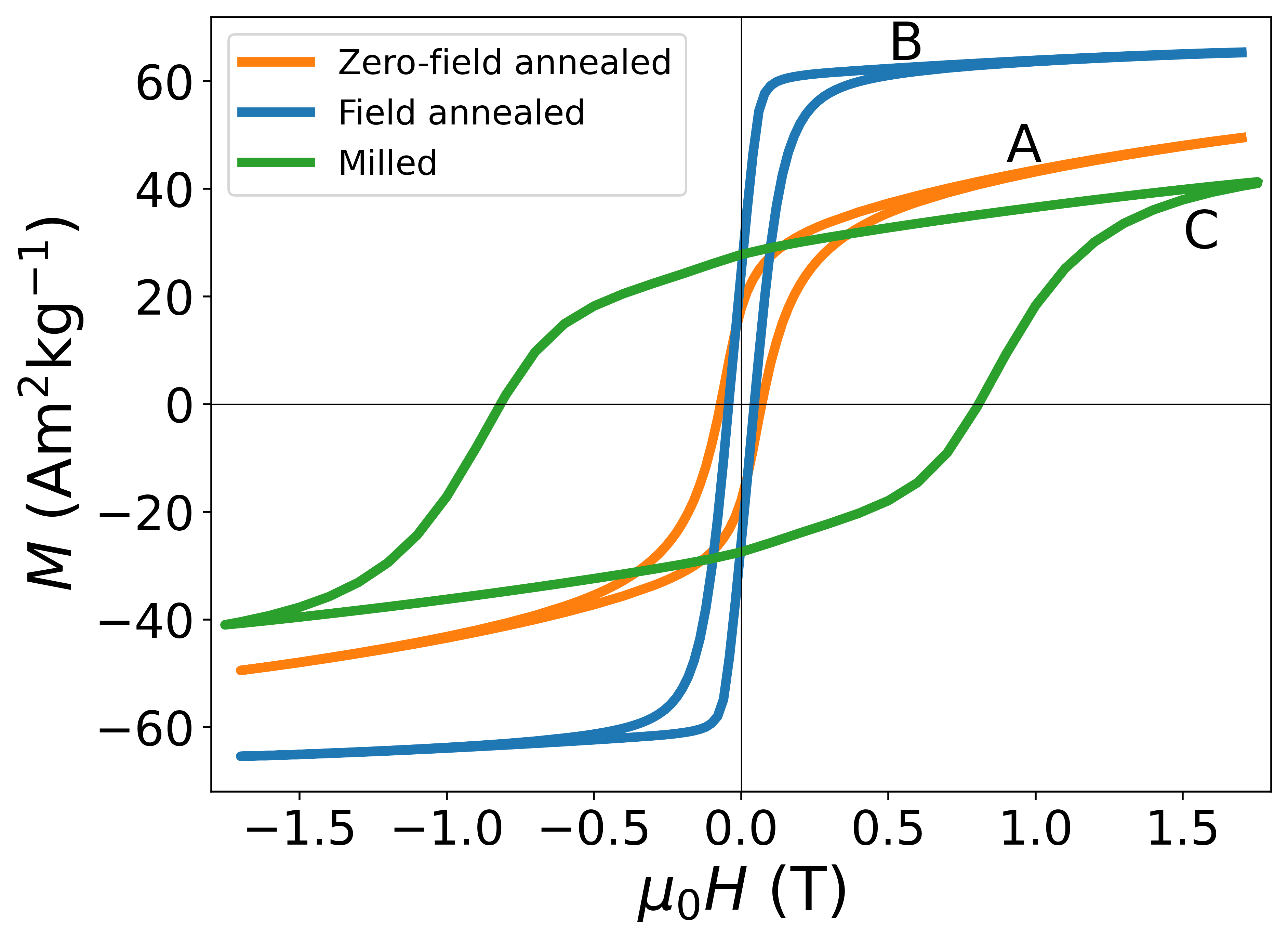}
\caption{\label{Fig1} Magnetization as a function of the applied magnetic field measured by vibrating sample magnetometry (VSM) for each type of polycrystalline MnBi sample. A: isotropic reference sample obtained by annealing without applied magnetic field. B: sample annealed under a static 1 T magnetic field. C: sample obtained by milling of the reference sample.}
\end{figure}

\begin{figure*}
\includegraphics[scale=0.39]{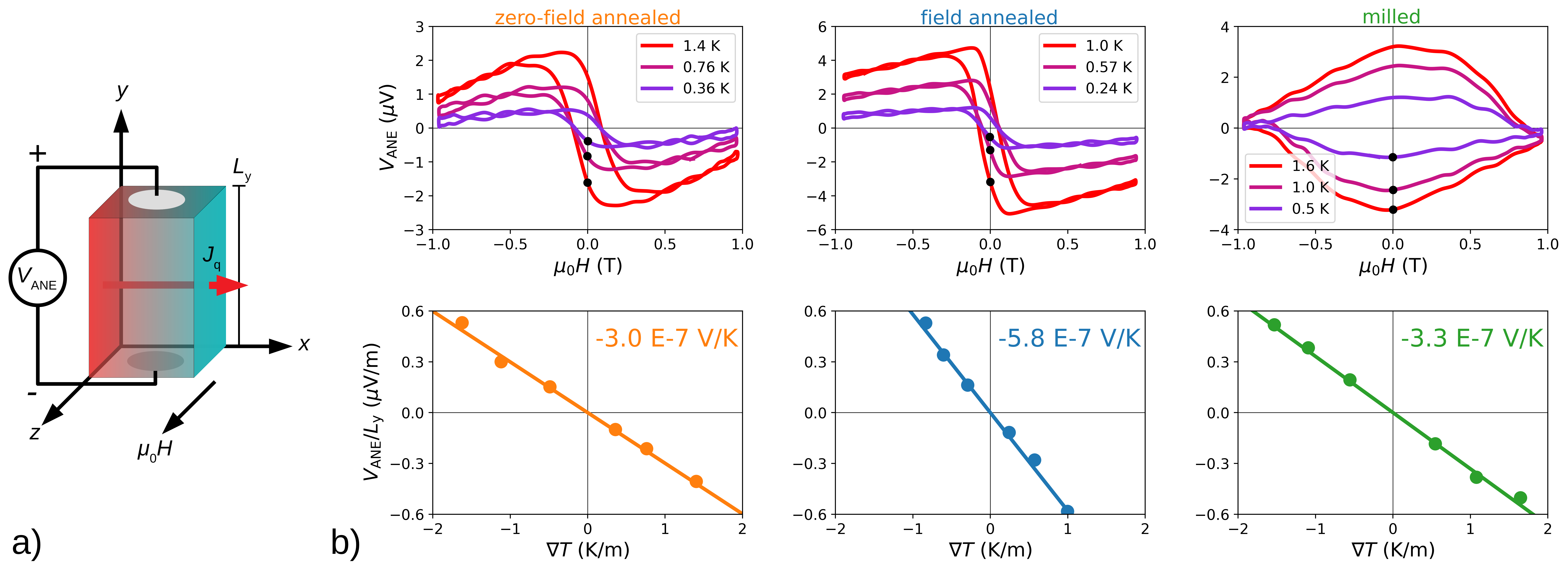}
\caption{\label{Fig2}a) Geometry of the ANE measurements: the sample is heated along the x-axis, magnetized along the z-axis and the $V_{\mathrm{ANE}}$ voltage is detected along the y-axis. The distance between electrodes is $L_{\mathrm{y}}$. b) Top: ANE voltages as a function of the applied magnetic field for three positive values of temperature difference. Bottom: gradients of ANE voltage as a function of the applied thermal gradient measured at zero magnetic field (corresponding to the values highlighted by the black dots on the top graphs). In the labels the values for the ANE thermopower are reported. Each column corresponds to the type of material: orange data-set for the zero-field annealed sample (A), blue data-set for the field annealed sample (B) and green data-set for the milled sample (C).}
\end{figure*}

\begin{figure}
\includegraphics[scale=0.35]{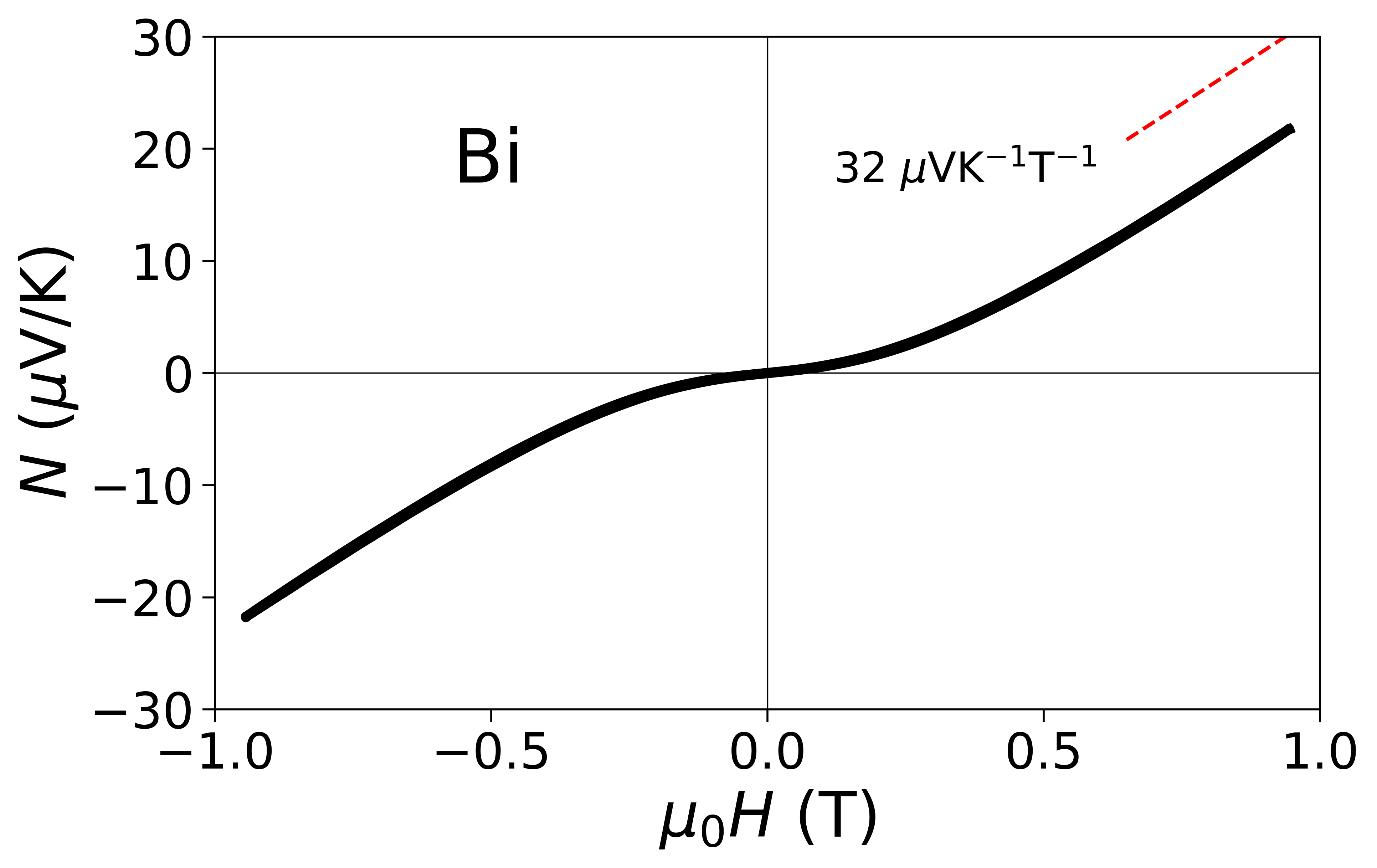}
\caption{\label{Fig3}Nernst thermopower of bismuth as a function of the applied magnetic field, normalized by the geometrical factors of the sample and by the thermal gradient. Experimental data are represented by the black line, while the red line is the reference slope for a Nernst coefficient of $32\cdot 10^{-6}$ VK$^{-1}$T$^{-1}$.}
\end{figure}

\begin{figure*}
\includegraphics[scale=0.5]{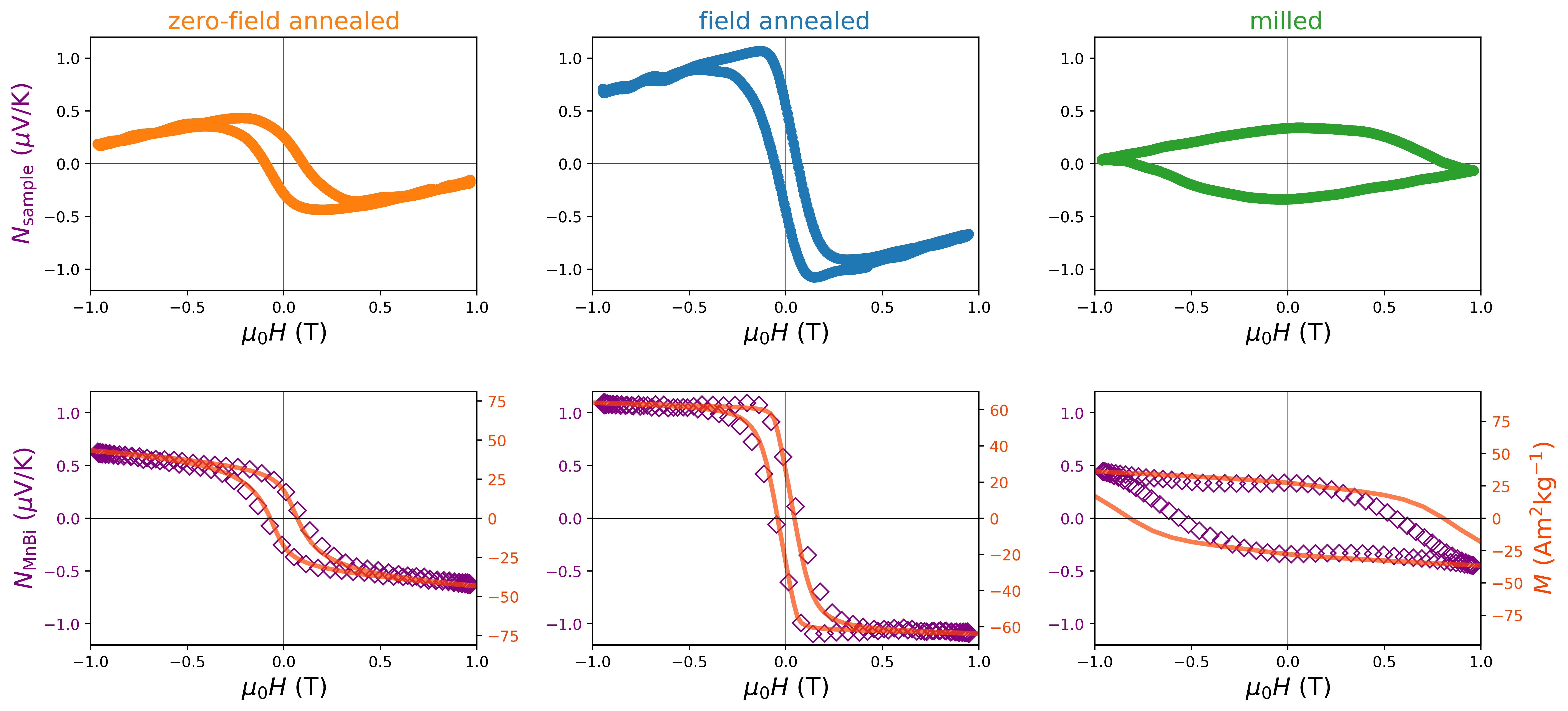}
\caption{\label{Fig4}Top: ANE thermopowers of MnBi samples as a function of the applied magnetic field, normalized by the geometrical factors. Bottom: comparison between VSM magnetization curves (orange line) and ANE thermopowers of MnBi (purple diamonds) after the compensation of the Nernst effect of bismuth, originating from the unreacted phase inside the samples.
The bismuth fractions needed to achieve the superimposition between the ANE and the VSM curves according to Eq. \ref{eq:2} is 2 \% for the zero-field annealed sample, 1.6 \% for the field annealed sample and 1.6 \% for the milled sample.}
\end{figure*}

The aim of the present study is the investigation of the ANE thermopower in polycrystalline MnBi samples prepared by powder metallurgy processes. Our study has the scope to unveil the gap in performance between ideal materials used for the study of fundamental phenomena and less ordered systems, with the advantage, in this second category, to exploit the hard magnetic properties of MnBi and have a device working without an external applied magnetic field.
Samples were prepared through a powder metallurgy route that consists in the annealing of manganese and bismuth powders, previously mixed and pressed to form a pellet. All the powders were handled in a nitrogen-filled glove box to avoid oxidation. The $\alpha$-MnBi phase is obtained by heating at 320 $^\circ$C for 1 hour in vacuum or inert atmosphere.
We prepared three types of samples, exploiting magnetic field annealing and the induction of defects by means of high-energy ball milling to tailor the microstructure and hence the magnetic properties of the samples\cite{curcio2015study}.
The hysteresis loops of the samples prepared according to these three procedures, measured by means of vibrating sample magnetometry (VSM), are shown in Fig. \ref{Fig1}.
Sample A, that can be considered as a reference, has been annealed in vacuum without applied magnetic field; this preparation favored the formation of randomly oriented and well developed MnBi grains. 
Sample B was annealed in nitrogen atmosphere under a static magnetic field of 1 T produced by a Halbach cylinder.
This procedure allows the development of a preferred orientation of the MnBi grains with their c-axis along the applied field direction. The resulting hysteresis cycle shows larger saturation and higher remanent magnetization compared to the reference sample. For sample C, we focused on the development of coercivity. In order to obtain a high value, we milled the material that was annealed in the same way as the sample A in a planetary ball mill. The milling process was for 1 h at 450 rpm in zirconia jar (ball-to-powder ratio $\sim$14) and afterwards we pressed the resulting powder in a new pellet. A smaller grain size and the presence of defects, representing pinning centres for domain wall motion, allowed to increase significantly the value of the coercive field \cite{curcio2015study}, as shown in Fig. \ref{Fig1}.
The transverse thermopower of the aforementioned samples was investigated by an experimental setup based on the heat flux method whose technical details and the calibration procedure are reported in previous works \cite{sola2017longitudinal,venkat2020measurement}. The experiment consists in the measurement of the transverse voltage response as a function of the heat flux traversing the sample. The latter quantity is monitored by means of calibrated Peltier cells while the whole system is kept under vacuum in order to avoid uncontrolled heat leakages. This procedure allows to disregard the role of the thermal resistances between the sample and the thermal sensor, that usually affects the direct measurement of the temperature difference. The value of the ANE thermopower $N$ has been experimentally derived as:
\begin{equation}
N=\frac{V_{\mathrm{ANE}}/L_{\mathrm{y}}}{-j_{\mathrm{q}}/k} \tag{1}\label{eq:1}
\end{equation}
where $V_{\mathrm{ANE}}$ is the measured voltage drop, $L_{\mathrm{y}}$ is the distance between the electrodes, $j_{\mathrm{q}}$ is the heat current density measured by the calibrated Peltier cells sensors and $k$ is the thermal conductivity of the material under investigation whose value, for our study, has been approximated to the one of pure bismuth that is $7.9$ WK$^{-1}$m$^{-1}$. A scheme of the measurement configuration is shown is Fig. \ref{Fig2} a. 
The magnetic field is perpendicular to the directions of the voltage drop $V_{\mathrm{ANE}}$ and the one of the heat current density $j_{\mathrm{q}}$.
An electromagnet was used to apply the magnetic field alternating between $\pm 1$ T and measured by a calibrated Hall probe. We compensated the induced voltage which constitutes a spurious component of the measured voltage drop $V_{\mathrm{ANE}}$; this component arose from the variation of the magnetic flux density across the electric wiring of the experiment. For each sample mounting inside the measurement system, the determination of these induced voltages was performed by applying the same varying magnetic field in absence of heat fluxes.
The results for the three types of MnBi samples are shown in Fig. \ref{Fig2} b.
The ANE voltages $V_{\mathrm{ANE}}$ as a function of the applied magnetic field are shown in the top row of Fig. \ref{Fig2} b. Here we represented only the values of $V_{\mathrm{ANE}}$ obtained with positive temperature differences, as an example of the measurement. The corresponding values of temperature difference are reported in the labels and have been derived as $\Delta T = -j_{\mathrm{q}}L_{\mathrm{x}}/k$. The measured voltage $V_{\mathrm{ANE}}$ follows an hysteresis loop whose magnitude increases along the vertical axis for higher temperature differences.
The values of $V_{\mathrm{ANE}}$ at zero applied field, highlighted by black dots on the hysteresis loops in the top row of Fig. \ref{Fig2} b, correspond to a Nernst effect driven by the remanent magnetization of the sample. 
These values, normalized by the geometrical factors of the samples, are used to represent $V_{\mathrm{ANE}}/L_{\mathrm{y}}$ at remanence as a function of the thermal gradients, shown in the bottom row of Fig. \ref{Fig2} b. Our system is able to produce heat currents $j_{\mathrm{q}}$ in both positive and negative directions along $x$; this allows to test the signs of the ANE voltages according to the signs of $j_{\mathrm{q}}$.
The slopes in Fig. \ref{Fig2} b are the values of ANE thermopower at remanence. Sample B, prepared under magnetic field annealing (blue set) exhibits the largest value, $N = -5.8 \cdot 10^{-7}$ V/K.
The curves shown in the top row of Fig. \ref{Fig2} b resemble the VSM measurements in Fig. \ref{Fig1}: particularly, sample B and C exhibit high remanent magnetization and high coercivity, respectively, also in their ANE characteristics. 
A decrease of the $V_{\mathrm{ANE}}$ for increasing applied magnetic fields is found in all samples; the interpretation of this phenomenon is the presence of unreacted bismuth, a source of transverse thermopower whose sign is opposite to the one of the MnBi ANE, as shown in Fig. \ref{Fig3}. 
Data in Fig. \ref{Fig3} are the measured Nernst thermopower of pure bismuth pellet obtained with the same experimental setup and in the same geometrical configuration used for the MnBi samples. The value of the Nernst coefficient of bismuth, $32\cdot 10^{-6}$ VK$^{-1}$T$^{-1}$, has been derived by the plateau of the the derivative of the curve of Fig. \ref{Fig3}. The red line is a guide for the eye whose slope is equal to the Nernst coefficient of bismuth and therefore parallel to the thermopower vs. field curve in the range above 1 T.
The three main features of the measured Nernst effect of bismuth are the high magnitude, the sign, opposite to the one of the MnBi ANE thermopower, and the behaviour as function of the applied magnetic field, that is non-linear but exhibits no hysteresis.
With this information it is possible to analyze the MnBi ANE measured thermopower by taking into account the presence of the bismuth Nernst voltage and considering the total measured voltage as:
\begin{equation}
V_{\mathrm{tot}}=(1-x_{\mathrm{Bi}})V_{\mathrm{MnBi}}+x_{\mathrm{Bi}}V_{\mathrm{Bi}} \tag{2}\label{eq:2}
\end{equation}
where $x_{\mathrm{Bi}}$ is the unreacted bismuth linear phase fraction inside the sample, $V_{\mathrm{Bi}}$ is the Nernst voltage contribution from bismuth and $V_{\mathrm{MnBi}}$ is the sought ANE voltage from MnBi. The criterion used to choose a value of $x_{\mathrm{Bi}}$ is to make the ANE loop follow the VSM loop for each type of material, by using the data reported in Fig. \ref{Fig3} for the $V_{\mathrm{Bi}}$. According to this calculation, it is possible to obtain the MnBi ANE voltage $V_{\mathrm{MnBi}}$ extrapolated to the absence of bismuth. The bottom row of Fig. \ref{Fig4} shows the same VSM data of Fig. \ref{Fig1} (orange lines) superimposed to the MnBi thermopowers (purple points) after the compensation of the Nernst contribution from bismuth, for the three types of materials. On the top row Fig. \ref{Fig4} are reported also the measured thermopowers from the samples, before the bismuth compensation.
The results of the MnBi ANE thermopowers at remanence together with the MnBi ANE thermopowers at saturating field after the bismuth compensation are summarized in Table \ref{tab:table1}, for the three types of materials.
\begin{table}
\caption{\label{tab:table1} ANE thermopowers of the materials under test}
\begin{ruledtabular}
\begin{tabular}{ccc}
MnBi Sample&ANE at 0 T&ANE at 1 T\\
      &$\mu$VK$^{-1}$ & $\mu$VK$^{-1}$\\
\hline
Zero-field annealed& -0.30 & -0.63\footnote{ANE thermopower for the MnBi after the bismuth compensation, obtained from the data reported on the bottom row of Fig. \ref{Fig4}} \\
Field annealed& -0.58 & -1.1$^{\text{a}}$ \\
Milled& -0.33 & -0.45$^{\text{a}}$ \\
\end{tabular}
\end{ruledtabular}
\end{table}
In the case of the field-annealed material (sample B), the value of ANE thermopower at 1 T reaches -1.1 $\mu$V/K, that is half of the value reported at room temperature for single crystal \cite{he2021large}. For the other types of materials (sample A and C) we observed lower ANE thermopowers. This is compatible with the anisotropic nature of the ANE, whose magnitude depends on the direction of the magnetization and of the c axis of the MnBi crystals.
We have further checked the value of the unreacted volume bismuth fraction by differential scanning calorimetry (DSC) measurements. The volume percentages of bismuth that we obtained were three to ten times bigger than the percentages needed to achieve the superimposition between the ANE thermopower curves and the magnetization curves of Fig. \ref{Fig4}, bottom. However, it is reasonable that the MnBi and the bismuth phases do not grow with the same shapes since the residual bismuth is presumably an intergranular phase that surrounds the MnBi crystals. 
With this model of the microstructure it is possible to evaluate the linear phase fraction $x_{\mathrm{Bi}}$ as the amount of phase crossed by a line oriented along an arbitrary direction inside the MnBi sample. 
The result is compatible to the ones used to meet the condition of the superimposition between magnetization curves and ANE thermopower curves.

In conclusion, we demonstrated the thermomagnetic characteristics of polycrystalline MnBi samples prepared by means of powder metallurgy followed by field annealing. This method permits an easy manipulation of the magnetic properties at the preparation stage and its direct effect on the ANE thermopower. This work paves the way towards the optimization of the properties of a transverse thermoelectric material such as MnBi: among the requirements, one of the most important is the high value of the remanent magnetization. Moreover, it is important to take into account the different contributions to the thermopower from materials whose thermomagnetic characteristics are different, like in the case of bismuth inside the MnBi samples. The optimization of these parameters allows the characteristics of polycrystalline samples to be comparable to the ones obtained from ideal and more difficult-to-produce materials, such as single crystals.
This represents a step forward in the applicability of thermomagnetic effects in the field of thermoelectric coolers and energy harvesting devices.

\nocite{*}
\bibliography{aipsamp}

\end{document}